\documentclass[twocolumn,english,superscriptaddress,floatfix,longbibliography]{revtex4-2}

\usepackage[english]{babel}
\usepackage{amsmath}
\usepackage{units}
\usepackage{upgreek}
\usepackage[colorlinks=true,urlcolor=blue,linkcolor=black,citecolor=black]{hyperref}
\usepackage{orcidlink}

\begin{document}
\title{A Low-Temperature Tunable Microcavity featuring High Passive Stability and Microwave Integration}
\author{Yanik Herrmann\orcidlink{0000-0003-0545-7002}}
\thanks{These authors contributed equally to this work.}
\author{Julius Fischer\orcidlink{0000-0003-2219-091X}}
\thanks{These authors contributed equally to this work.}
\author{Stijn Scheijen\orcidlink{0009-0009-4092-4674}}
\author{Cornelis F. J. Wolfs\orcidlink{0009-0007-1776-3441}}
\author{Julia M. Brevoord\orcidlink{0000-0002-8801-9616}}
\author{Colin Sauerzapf\orcidlink{0009-0004-7655-9289}}
\altaffiliation[Present address: ]{3rd Institute of Physics and Research Center SCoPE, University of Stuttgart, 70049 Stuttgart, Germany}
\author{Leonardo G. C. Wienhoven\orcidlink{0009-0009-7745-3765}}
\author{Laurens J. Feije\orcidlink{0009-0009-6909-9145}}
\affiliation{QuTech and Kavli Institute of Nanoscience, Delft University of Technology, P.O. Box 5046, 2600, GA Delft, The Netherlands}
\author{Martin Eschen\orcidlink{0009-0000-7336-6132}}
\affiliation{QuTech and Kavli Institute of Nanoscience, Delft University of Technology, P.O. Box 5046, 2600, GA Delft, The Netherlands}
\affiliation{Netherlands Organisation for Applied Scientific Research (TNO), P.O. Box 155, 2600 AD Delft, The Netherlands}
\author{Maximilian Ruf\orcidlink{0000-0001-9116-6214}}
\altaffiliation[Present address: ]{SandboxAQ, Palo Alto, California, USA}
\author{Matthew J. Weaver\orcidlink{0000-0002-7093-4058}}
\altaffiliation[Present address: ]{QphoX B.V., Elektronicaweg 10, 2628 XG Delft, The Netherlands}
\author{Ronald Hanson\orcidlink{0000-0001-8938-2137}}
\email{R.Hanson@tudelft.nl}
\affiliation{QuTech and Kavli Institute of Nanoscience, Delft University of Technology, P.O. Box 5046, 2600, GA Delft, The Netherlands}

\date{\today}

\begin{abstract}
Open microcavities offer great potential for the exploration and utilization of efficient spin-photon interfaces with Purcell-enhanced quantum emitters thanks to their large spectral and spatial tunability combined with high versatility of sample integration. However, a major challenge for this platform is the sensitivity to cavity length fluctuations in the cryogenic environment, which leads to cavity resonance frequency variations and thereby a lowered averaged Purcell enhancement. This work presents a closed-cycle cryogenic fiber-based microcavity setup, which is in particular designed for a low passive vibration level, while still providing large tunability and flexibility in fiber and sample integration, and high photon collection efficiency from the cavity mode. At temperatures below $\unit[10]{}$ Kelvin, a stability level of around $\unit[25]{}$ picometer is reproducibly achieved in different setup configurations, including the extension with microwave control for manipulating the spin of cavity-coupled quantum emitters, enabling a bright photonic interface with optically active qubits.
\end{abstract}

\maketitle

%\begin{quotation}
%``Achtung! Now there are only two possible outcomes: Either it works or it doesn't work.'' %- Luke in ``Jim Button and Luke the Engine Driver'' by Michael Ende.
%\end{quotation}

\section{Microcavities with Single Quantum Emitters}
\label{sec:intro}

\begin{figure}[ht]
    \centering
    \includegraphics[width=\linewidth]{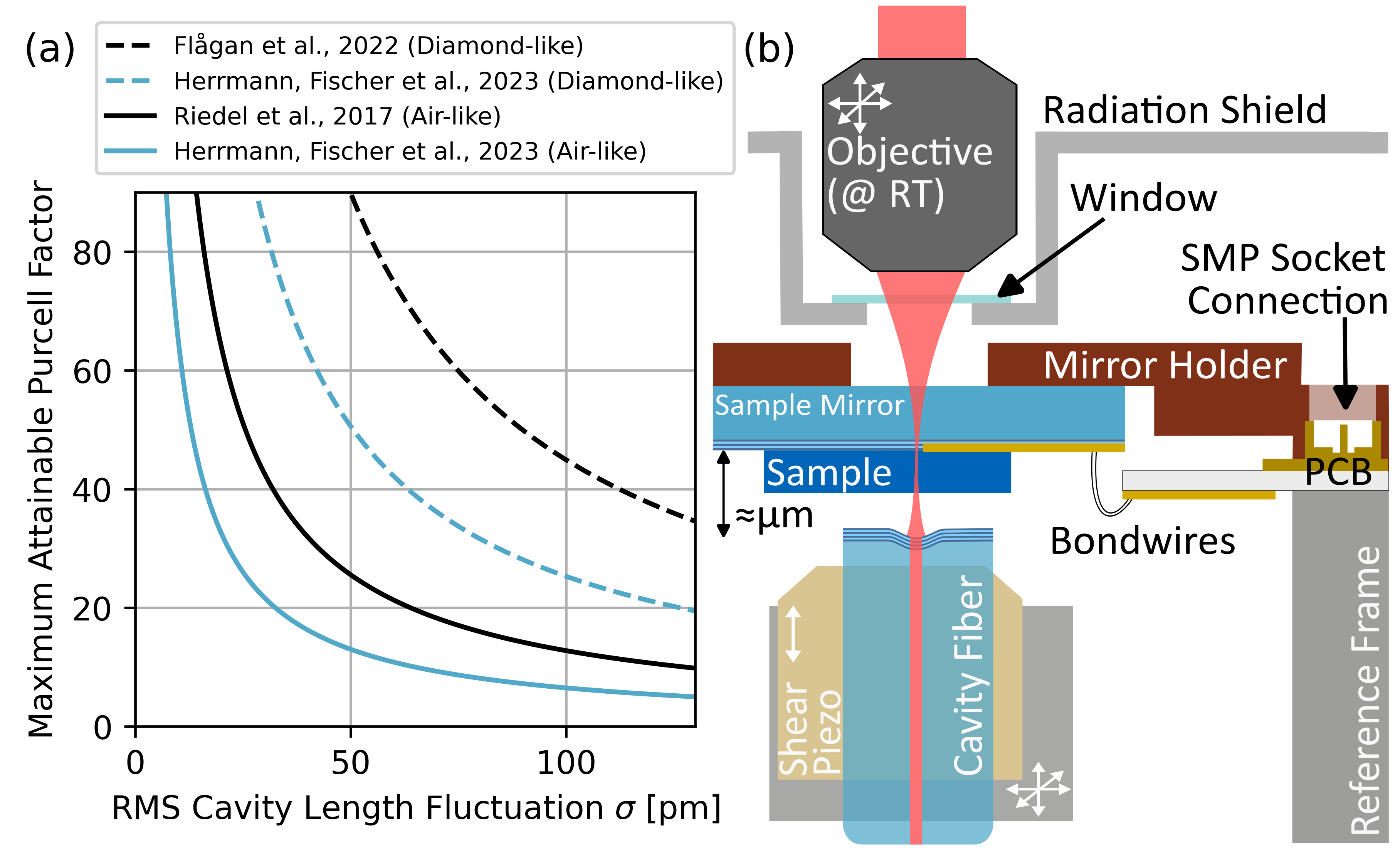}
    \caption{(a) Maximum attainable Purcell factor depending on RMS cavity length fluctuations. The black lines show simulations for a microcavity with an around $\unit[0.8]{\upmu m}$ thin diamond membrane, achieved in Ref. \cite{riedel_deterministic_2017} for an air-like mode (mode volume $V=\unit[9]{\lambda^3}$, cavity dispersion $s=\unit[139]{MHz/pm}$ for $\unit[637]{nm}$) and in Ref. \cite{flagan_diamond-confined_2022} for a diamond-like mode (mode volume $V=\unit[4]{\lambda^3}$, cavity dispersion $s=\unit[89]{MHz/pm}$ for $\unit[637]{nm}$). The blue lines show simulations for an around $\unit[3.7]{\upmu m}$ thin diamond membrane, achieved in Ref. \cite{herrmann_coherent_2024} for an air-like mode (mode volume $V=\unit[55]{\lambda^3}$, cavity dispersion $s=\unit[46]{MHz/pm}$ for $\unit[619]{nm}$) and for the diamond-like case (mode volume $V=\unit[31]{\lambda^3}$, cavity dispersion $s=\unit[21]{MHz/pm}$ for $\unit[619]{nm}$). For a cavity without the diamond membrane, the maximum attainable Purcell factor follows approximately the air-like modes. (b) Sketch of the experimental setup design. The microcavity is composed of a flat sample mirror and a curved microscopic mirror, fabricated on the tip of an optical fiber. Optical access to the sample mirror side of the cavity is provided with a movable objective. The optical cavity axis is changed by moving the cavity fiber and the outcoupling objective over the sample mirror. The objective is kept at room temperature and thermally shielded by an aluminum cone with a thin window for optical access. The sample mirror has embedded gold striplines, which are connected over a support printed circuit board (PCB), to deliver microwaves close to the cavity spot.}
    \label{fig:schematic}
\end{figure}

The strength of the light-matter interaction is a key parameter for the realization of efficient interfaces between single quantum emitters and optical photons. Optically active quantum emitters are interesting testbeds for quantum science and are promising candidates for realizing stationary qubits \cite{awschalom_quantum_2018, atature_material_2018, wolfowicz_quantum_2021}, but the bare coupling to optical photons in bulk materials is generally weak. This coupling can be greatly improved by integrating the emitter into an optical resonator \cite{vahala_optical_2003, reiserer_cavity-based_2015, janitz_cavity_2020, ruf_quantum_2021}. In the simplest form, two highly reflective mirrors facing each other can be used to realize such a cavity. The cavity can be utilized to efficiently couple quantum emitters to an optical photon mode thereby selectively enhancing its emission, known as the Purcell effect. When used with solid-state host materials, a hemispherical plano-concave cavity geometry with a low radius of curvature and a micrometer short length can be used to minimize the mode volume. Such open optical microcavities have proven to be a versatile tool in quantum optics, due to their high spectral and spatial tunability and their compatibility with a broad spectrum of quantum systems. The cavity design allows to incorporate quantum emitters in various ways: quantum dots can be directly grown on top of the mirror \cite{muller_coupling_2009,di_controlling_2012,najer_gated_2019,antoniadis_cavity-enhanced_2023} or emitters in layered materials \cite{vogl_compact_2019,vadia_open-cavity_2021,tan_bose_2023} can be bonded onto the mirror surface. Furthermore, the spatial tunability can be used to optimize on localized emitters on the mirror like carbon nanotubes \cite{hummer_cavity-enhanced_2016,borel_telecom_2023}, rare-earth ions in nanoparticles \cite{casabone_dynamic_2021} or color centers in nanodiamonds \cite{albrecht_coupling_2013,johnson_tunable_2015,benedikter_cavity-enhanced_2017,bayer_optical_2023}. And (sub) micrometer thin membranes can integrate color centers in host materials like diamond \cite{yurgens_cavity-assisted_2024-1,berghaus_cavity-enhanced_2024,zifkin_lifetime_2024}, silicon carbide \cite{lukin_integrated_2020, heiler_spectral_2024}, yttrium orthosilicate \cite{merkel_coherent_2020} or organic crystals \cite{wang_turning_2019} into the cavity. This approach is especially advantageous for emitters like the Nitrogen-Vacancy center in diamond, which are notoriously hard to integrate coherently into nanostructures due to the presence of a permanent electric dipole, but maintain good optical properties in micrometer-thin membranes \cite{ruf_optically_2019}.\\
The optical access of the microcavity can conveniently benefit from the realization of the mirrors on the tip of optical fibers \cite{hunger_fiber_2010,hunger_laser_2012}, enabling direct fiber coupling. Furthermore, high cavity quality factors can be realized by commercially available mirror coatings, including the purposely design of symmetric or single-sided cavities.\\
The high spectral tunability makes the cavity sensitive to fluctuations in length, leading to variations in the resonance frequency. These perturbations lower the effective coupling of the emitter to the cavity and lead to a reduced averaged Purcell enhancement. This can be expressed with the spectral overlap of the resonance frequency of the emitter and the cavity frequency \cite{van_dam_optimal_2018}. Assuming a Gaussian distribution of the varying cavity length \cite{ruf_resonant_2021}, we can find a bound for the maximum attainable (effective) Purcell factor
\begin{equation}
    F_{P,max} = \frac{3}{4 \pi^2} \left(\frac{c}{n \nu}\right)^3 \frac{1}{V} \sqrt{\frac{\pi}{2(s\sigma)^2}} \frac{\nu}{2},
\label{equ:F_P}
\end{equation}
which depends on the cavity resonance frequency $\nu$, mode volume $V$, mode dispersion slope $s$, and root mean square (RMS) cavity length fluctuations $\sigma$ with the refractive index in diamond $n$ and the speed of light $c$. The derivation of equation \eqref{equ:F_P} can be found in Appendix~\ref{app:derivation}. For a hybrid cavity consisting of an air part and a diamond membrane, certain modes of a fixed frequency become more air- or diamond-like \cite{janitz_fabry-perot_2015}. The mode type depends only on the diamond thickness. Due to a smaller slope in the cavity dispersion relation, diamond-like modes are less sensitive to vibrations \cite{van_dam_optimal_2018}. Note that for a given membrane thickness and air length, the cavity mode dispersion slope $s$ can be calculated with an analytic expression \cite{van_dam_optimal_2018}. Figure~\ref{fig:schematic}~(a) shows the maximum attainable Purcell factor depending on the RMS cavity length fluctuations for different cavity parameters realized in other works. To reach a large Purcell enhancement, a vibration level on the order of tens of picometer RMS vibrations is required. This poses a technical challenge, as most quantum systems need to be operated in high vacuum at cryogenic temperatures. For a closed-cycle cryostat configuration, which enables continuous operation and reduced experimental overhead, the vicinity of a running cryostat cold head inevitably introduces vibrations.\\
Setups with high stability levels have been reported in quiet helium bath cryostats \cite{greuter_small_2014} or with limited spatial control in closed-cycle cryostats \cite{merkel_coherent_2020}. Recently, systems with large spatial tunability and a passive stability levels at the tens of picometer level have been reported in closed-cycle cryostats \cite{fontana_mechanically_2021,ruelle_tunable_2022,pallmann_highly_2023,fisicaro_active_2024}.\\
In the following, we present a low-temperature fiber-based Fabry-P\'{e}rot microcavity setup inside a closed-cycle cryostat, which is designed to maintain a high passive stability level. We present the operation of the cavity together with an analysis of the passive stability. We showcase the functionality of this tunable platform by demonstrating the coupling of two different diamond color centers, the Tin-Vacancy center and the Nitrogen-Vacancy center, to the cavity at temperatures below $\unit[10]{K}$. Furthermore, we show that the integration of microwave delivery lines into our system, to address the spin states of cavity-enhanced Nitrogen-Vacancy centers, does not change the passive stability level.

\section{Setup Design}
\label{sec:setup}

A schematic drawing of the cavity geometry and the setup design is shown in Fig.~\ref{fig:schematic}~(b). The microcavity is mounted in an optical cryostat (Montana~Instruments~HILA) with an off-table cold head design and a base temperature of $\unit[5.6]{K}$. The cold head is a two-stage Gifford-McMahon cryocooler, which is suspended inside a rack-mounted cooling tower assembly, hovering above the optical table. To minimize the vibration transfer, the cold head tower is connected over a loose bellow for vacuum and flexible braiding for cooling to the main chamber (see photograph in Appendix Fig.~\ref{fig:chamber}).\\
A computer-aided design (CAD) software drawing of the cryostat baseplate with the microcavity insert is shown in Fig.~\ref{fig:drawing}~(a) and a photograph in Appendix Fig.~\ref{fig:insert_chamber}~(a). The baseplate is integrated into an auto-leveling floating stage acting as a low-pass filter for mechanical vibrations with a cutoff frequency of about $\unit[1]{Hz}$. The baseplate is separated into three parts: an outer ring, kept at room temperature to mount the objective cone. An intermediate ring, which is connected to stage one of the cold head, reaching about $\unit[90]{K}$. This ring is used for connecting the electrical cables and mounting the radiation shield separating microcavity and room temperature objective. The microcavity insert is placed on the central part, connected to stage two of the cold head reaching the base temperature. The cryostat chamber (photograph shown in Appendix Fig.~\ref{fig:insert_chamber}~(b)) is constantly pumped to a pressure in the range of $\unit[1 \times 10^{-6}]{mBar}$, by an integrated turbopump and a differentially pumped outer vacuum.\\
The flexibility of moving the fiber over a millimeter-large range comes usually with the drawback of lower mechanical resonance frequencies, making the positioning system more susceptible to vibrations \cite{yong_invited_2012}. With a microcavity, where the fiber is placed on an orthogonal XYZ nanopositioning stage (JPE~CS021), we measure a vibration level of about $\unit[0.5]{nm}$ directly on the baseplate at room temperature. A possible explanation for this is the low mechanical resonance frequencies of the nanopositioner stack (typically on the order of several $\unit[100]{Hz}$), which are less suppressed by the isolation system.\\
To improve the vibration level, a different positioning system with a high internal resonance frequency is chosen. This is provided by the cryo positioning stage high resonance (JPE~CPSHR1-a) placed on a passive vibration isolator (JPE~CVIP1). The cross section through the stage is shown in Fig.~\ref{fig:drawing}~(b). An optimized stiffness due to parallel kinematics results in a high resonance frequency of about $\unit[4]{kHz}$ along the cavity axis \cite{jpe_cpshr1_datasheet_2024}. Three linear actuators (JPE~CLA2201) move the fiber in-situ in a tripod configuration over a range of more than $\unit[2]{mm}$ laterally and $\unit[1]{mm}$ along the cavity length axis, with a minimum step size of $\unit[1]{nm}$ at low temperatures. Additionally, each tripod axis is equipped with a fine scanning piezo element, with sub-nanometer resolution. This allows continuous cavity length detuning with frequencies up to $\unit[5]{kHz}$ over a range of up to $\unit[0.5]{\upmu m}$ at low temperatures (all specifications of the CPSHR1-a stage can be found online \cite{jpe_cpshr1_website_2024}).\\
We employ two different ways of mounting the fiber, displayed in Fig.~\ref{fig:drawing}~(d): directly glued on a titanium holder or onto a shear piezo element featuring a high resonance frequency (Noliac~CSAP02). The latter enables high-bandwidth fine control of the cavity length, allowing for active cavity length stabilization \cite{janitz_high_2017}. In both settings, we use a two-component epoxy (Stycast~2850FT with catalyst LV~24) to glue the fiber. To guide the cavity fiber into the vacuum chamber, we make use of a $\unit[1/8]{in}$ stainless steel tube fitting (Swagelok male connector) with a custom-made Teflon ferrule. This ferrule has a borehole of $\unit[0.17]{mm}$ to fit the cavity fiber (Coherent FUD-4519 S630-P single-mode fiber) with a polyimide protection coating. Furthermore, we use a $\unit[0.25]{mm}$ thin polytetrafluoroethylene~(PTFE) film (Reichelt Chemietechnik), cut to pieces of approximately $\unit[5]{mm}$ width and $\unit[25]{mm}$ length to further damp the fiber movement. This film connects the fiber mount and the reference frame (not shown in Fig.~\ref{fig:drawing}, see photograph in Appendix~Fig.~\ref{fig:fiber_sample}~(a)). We see that the additional damping by the film can improve the vibration level.\\
The sample mirror is mounted on top of the positioning system, separated by $\unit[> 100]{\upmu m}$ to the fiber. To ensure a good thermalization of the sample mirror to the baseplate, three flexible links made out of highly conductive copper (Montana~Instruments custom-design) are used (not shown in the drawing, see Appendix Fig.~\ref{fig:insert_chamber}~(a)). Each link is composed of multiple flat braids to ensure a good thermal connection in combination with a low transfer of vibrations from the baseplate to the cavity. This keeps the temperature difference between the mirror holder and baseplate to about $\unit[0.2]{K}$. On the mirror holder, we monitor the temperature with a temperature sensor (Lake~Shore Cryotronics~Cernox~1050). To further ensure that the sample mirror reaches low temperature, a $\unit[0.2]{mm}$ thin window (Edmond~Optics Ultra-Thin~N-BK7, with anti-reflection coating for $\unit[425]{nm}$ to $\unit[675]{nm}$) glued into the radiation shield is used. This reduces considerably the heat load stemming from the black-body radiation of the room temperature objective that is about $\unit[4]{mm}$ away. The sample mirror is glued with silver conductive paint to the mirror holder, shown in Fig.~\ref{fig:drawing}~(c). In a test cooldown, we measure the temperature directly on the center of the sample mirror with a temperature sensor (Lake~Shore Cryotronics~Cernox~Thin~Film~RTD), glued onto it with GE low-temperature varnish. Without the window, a temperature of around $\unit[20]{K}$ on the sample mirror is measured, while a temperature of about $\unit[8]{K}$ is reached with the window. This temperature is consistent with the measured linewidth of cavity-integrated diamond Tin-Vacancy centers in previous work \cite{herrmann_coherent_2024}.\\
To control the spin state of quantum emitters inside the cavity, microwaves are delivered to the cavity spot with gold striplines, embedded into the sample mirror \cite{bogdanovic_robust_2017}. The striplines are interfaced with $\unit[25]{\upmu m}$ thin standard wire bonds to a support PCB, glued onto the cryostat insert (see Fig.~\ref{fig:drawing}~(b),(c) and Appendix Fig.~\ref{fig:fiber_sample}~(b)). The PCB is connected with flexible coax cables (Montana~Instruments LF-5 with SMP female ends) to semi-rigid coax cables (Montana~Instruments~C-20) on the baseplate, which are interfaced over a vacuum microwave feedthrough.\\
Optical access to the sample mirror side of the cavity is provided by a standard objective (Zeiss LD~EC~Epiplan-Neofluar, 100x~magnification, $\unit[0.75]{}$ numerical aperture, $\unit[4]{mm}$ working distance). The objective can be positioned in a tripod configuration with three linear actuators (Physik~Instrumente~Q-545), mounted on an encircling stainless steel cone shown in Appendix Fig.~\ref{fig:insert_chamber}~(b).

\onecolumngrid
\begin{center}
\begin{figure*}[ht]
    \centering
    \includegraphics[width=\linewidth]{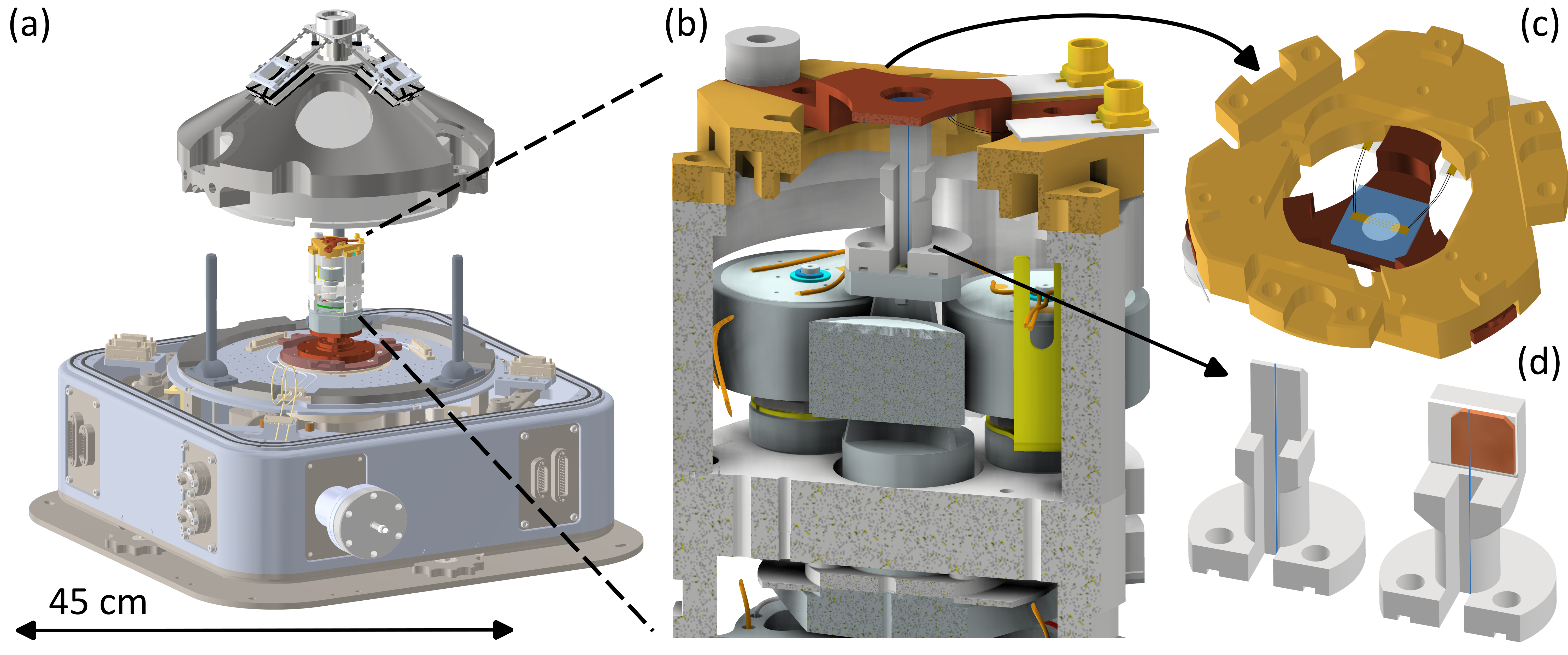}
    \caption{CAD drawing of the full setup with detailed individual parts. The drawing of the HILA cryostat in (a) is made by Montana Instruments and the drawings of the positioner stage in (a) and (b) and the adapter piece in (c) are from JPE, used with permission and available under Ref. \cite{jpe_cpshr1_website_2024}. The drawing of the linear actuators in (a) is made by Physik Instrumente, used with permission, and is available under Ref. \cite{pi_q_545_website_2024}. The drawing of the temperature sensor in (a) and (b) is made by Lakeshore, used with permission, and is available under Ref. \cite{lakeshore_website_2024}. (a) Floating cryostat baseplate with microcavity insert inside the vacuum chamber. The front part shows the fiber, microwave, and electrical feedthroughs. The insert is mounted on the copper-colored inner part of the baseplate, reaching a base temperature of $\unit[5.6]{K}$. Two D-Sub~19 feedthroughs on the baseplate are used to connect the actuators, piezo elements, and temperature sensor to the control and readout electronics. The middle ring of the baseplate mounts the radiation shield (at about $\unit[90]{K}$), while the outer part is used for the objective cone (at room temperature). (b) Cross section of the microcavity insert. The fiber mount is positioned by three linear actuators in a tripod configuration. The sample mirror is glued on a copper holder, which is directly connected with flexible thermal links to the baseplate (not shown here, see photograph of the setup in the Appendix Fig.~\ref{fig:insert_chamber}~(a)). The sample mirror has a thickness of about $\unit[0.5]{mm}$, and is placed at a distance of $\unit[> 100]{\upmu m}$ to the fiber, which can be compensated by the positioning system (a detailed sketch of the full microcavity insert can be found in Appendix Fig.~\ref{fig:positioning}). Each moving axis has an additional piezo element for fine positioning. We monitor the sample temperature with a sensor mounted on top of the microcavity insert. Furthermore, a support PCB is glued onto the stack to connect the mirror striplines to flexible coax cables. (c) Sample mirror on the copper holder with an adapter piece (JPE I2-CPSHR1) connecting to the reference frame. The gold striplines embedded into the mirror coating are connected with standard wire bonds to the support PCB. (d) Mounting of the cavity fiber: the fiber is glued either directly on a titanium mount (left) or onto a shear piezo element for fast fine control (right). The length of the fiber sticking out is kept as short as possible (between $\unit[100]{\upmu m}$ and $\unit[200]{\upmu m}$) to avoid possible resonances associated with the overhanging fiber tip \cite{janitz_high_2017}.}
    \label{fig:drawing}
\end{figure*}
\end{center}
\twocolumngrid

\section{Microcavity Operation}

\begin{figure}[ht]
    \centering
    \includegraphics[width=\linewidth]{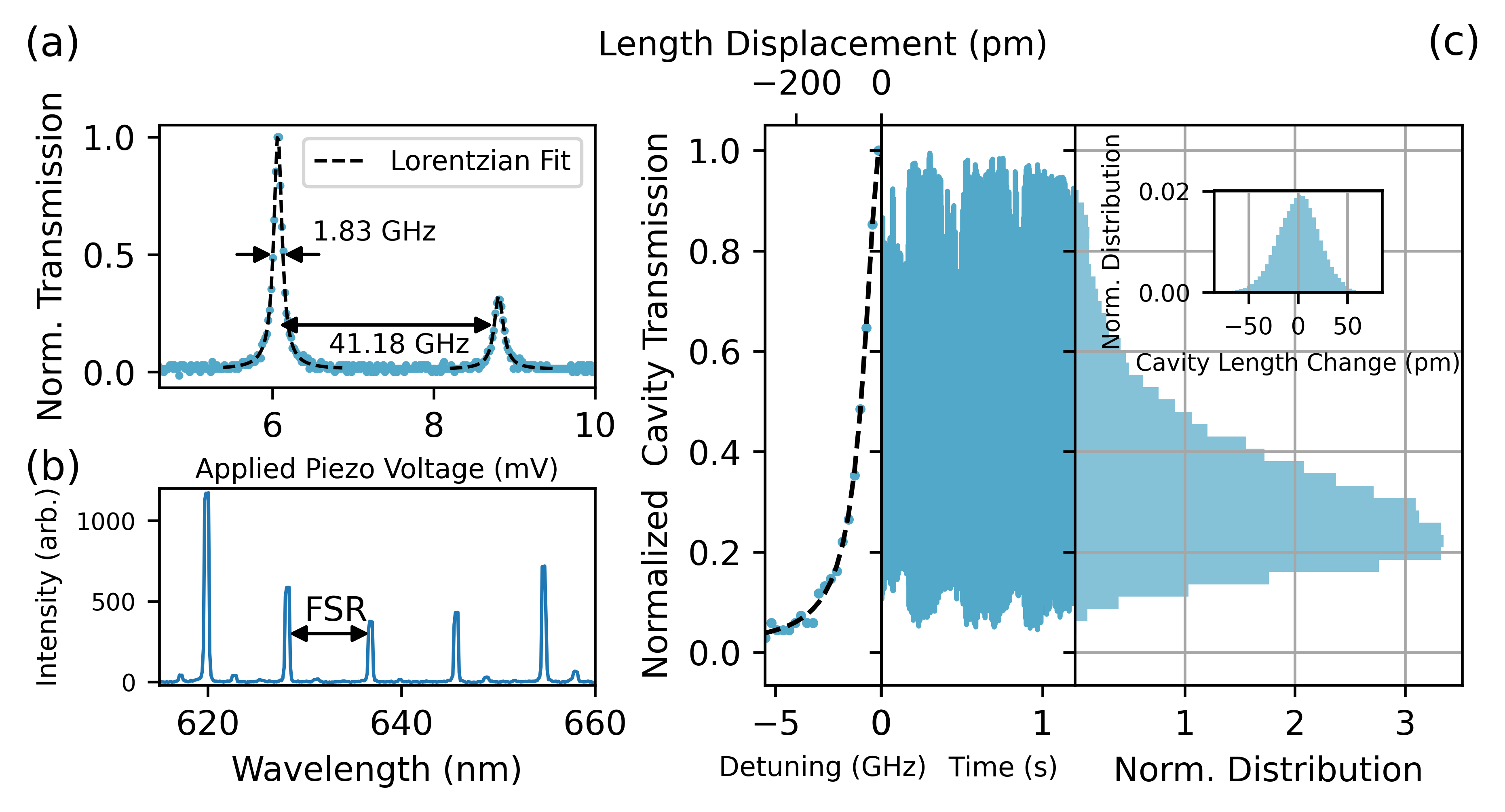}
    \caption{Characterizing the microcavity properties and the vibration level. (a) Cavity linewidth measurement with two detuned resonant lasers. The cavity transmission is probed by applying a voltage to one of the fine piezo elements to scan the cavity length (scan frequency of $\unit[2.33]{kHz}$). (b) Cavity length measurement by probing the transmission of a broadband supercontinuum white light source. The visible fundamental modes directly determine the free spectral range and the cavity length. (c) Exemplary cavity vibration measurement: the cavity transmission signal (center panel) is mapped on half of the Lorentzian cavity resonance (left panel). The spatial linewidth of the cavity allows to translate the transmission signal into a change in length. The right panel shows a histogram of the transmission for the full measurement ($\unit[10]{s}$). The insert shows the corresponding distribution of the cavity length values.}
    \label{fig:operation}
\end{figure}

The setup is controlled via a PC and the Python 3 framework QMI \cite{raa_qmi_2023}. We use a real-time microcontroller (J\"{a}ger~Computergesteuerte~Messtechnik Adwin~Pro~II) for analog voltage control of the fine piezo element offset. Cavity scans with a frequency of up to $\unit[5]{kHz}$ are accomplished with a signal generator, connected to the fine piezo via a bias-T. The signal is amplified (JPE Piezo Scanning Module) and filtered afterward with a home-built voltage-controlled, switchable lowpass filter (cutoff frequency in the range of $\unit[30]{Hz}$). Without filtering, we see that the noise from the amplifier can excite the resonance frequencies and increase the vibration level by a factor of up to two. The driving voltage of the linear actuators is also filtered and controlled by a high voltage amplifier (JPE~Cryo~Actuator~Driver~Module~2).\\
The microcavity is characterized by linewidth and length (or air length and membrane thickness). With these parameters, important properties like finesse, quality factor, and mode volume can be calculated, and the sensitivity to vibrations can be determined. To measure the linewidth, we probe the cavity in transmission with two resonant lasers (Newfocus~Velocity~TLB-6300-LN and Toptica~DL~Pro), which are frequency-stabilized to a reference wavemeter (High~Finesse~WS-U). The cavity transmission signal is measured with a free-space photodiode (Thorlabs~APD130A2). In Fig.~\ref{fig:operation}~(a) we scan the cavity resonance over the two laser frequencies by applying a voltage to one of the fine piezo elements moving the fiber. The difference in laser frequency is used to transform the scanning voltage into a change in cavity resonance frequency and to fit the two transmission peaks in the frequency domain. The use of two lasers with an arbitrary detuning allows to determine a narrow, but also broader ($\unit[> 10]{GHz}$) cavity linewidth. This contrasts the use of a single laser with sidebands imprinted by an electrooptic phase modulator \cite{ruf_resonant_2021}, in which case the range is limited by the frequency spacing of sidebands. To determine the cavity length, we probe the cavity in transmission with a supercontinuum white light source (NKT~Photonics~SC-450-2), filtered to $\unit[600]{nm}$ to $\unit[700]{nm}$, which fully covers the cavity stopband (Fig.~\ref{fig:operation}~(b)). The transmission signal is sent to a fiber-coupled spectrometer (Princeton~Instruments~SP-2500i). The cavity acts as a spectral filter, where fundamental modes appear as bright peaks in transmission. These modes are spaced by the free spectral range, which is directly related to the cavity length.\\
During cooling to the base temperature, the microcavity insert shrinks in height by about $\unit[550]{\upmu m}$ (measured by the difference in the focal position of the objective), and the cavity length is reduced by about $\unit[50]{\upmu m}$.\\
Imaging through the objective with an LED and camera provides accurate observation of the cavity fiber position and the sample. The high magnification of the objective makes it possible to move the fiber with micrometer precision laterally over the sample mirror.\\
The cavity length stability as the crucial system parameter needs to be studied for different system configurations. For a given cavity length, the measured spectral cavity linewidth can be translated into a spatial linewidth, which allows direct measurement of the cavity length fluctuations. We use the same method as used in earlier work \cite{ruf_resonant_2021}, where the cavity transmission signal is recorded and mapped to a change in cavity lengths (Fig.~\ref{fig:operation}~(c)). The length is monitored with the cavity transmission signal, measured on a photodiode ($\unit[50]{MHz}$ bandwidth) over a time of $\unit[10]{s}$. This time trace is used to calculate the RMS vibrations, and a Fourier transform gives insights into the corresponding mechanical resonance frequencies. The cavity is operated in the low Finesse regime ($\unit[1000]{}$ to $\unit[4000]{}$) to make sure that the transmission signal is fully caught on one half of the Lorentzian linewidth. Overshooting would lead to an underestimated cavity vibration level. To avoid this, we make sure that all points of the acquired length displacement trace follow an approximate Gaussian distribution (inset in Fig.~\ref{fig:operation}~(c)).

\section{Performance of System at Low Temperatures}

\begin{center}
\begin{figure*}[ht]
    \centering
    \includegraphics[width=\linewidth]{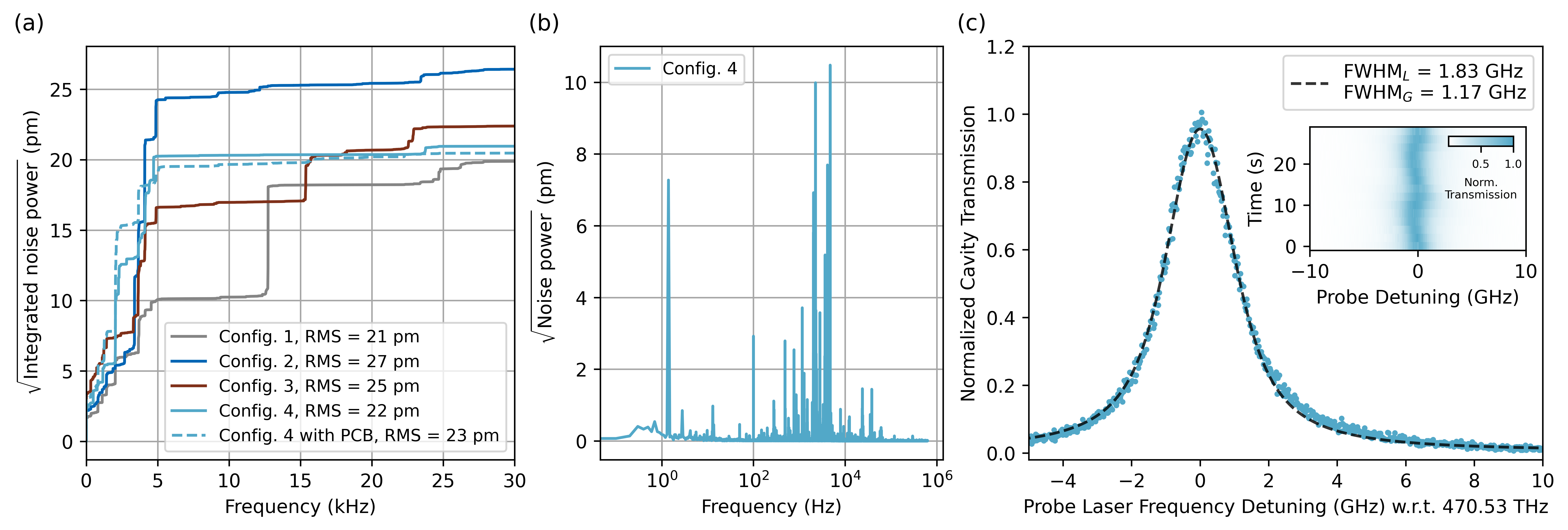}
    \caption{System performance at low temperatures for different setup configurations. (a) A comparable vibration level is reached for different settings: for the first three configurations, the fiber is mounted on a shear piezo element (right side of Fig.~2~(d)), while configuration four is without (left side of Fig.~2~(d)). All configurations use different cavity fibers and sample mirrors, configuration two was used in previous work with diamond Tin-Vacancy centers \cite{herrmann_coherent_2024}. The extended PCB for microwave control in configuration four does not influence the vibration level. (b) Resonance spectrum of configuration four, revealing a typical resonance at $\unit[1.4]{Hz}$ corresponding to the cold head cycle and resonances up to $\unit[5]{kHz}$ attributed to the fiber positioning system. (c) Resonant cavity transmission scan for configuration four. The cavity is probed with a scan speed of about $\unit[10]{GHz/s}$. The data is fitted with a Voigt profile, in which the Lorentzian part is fixed to the cavity linewidth. The Gaussian component can be used to calculate the vibration level. The inset shows $\unit[15]{}$ consecutive scans.} 
    \label{fig:performance}
\end{figure*}
\end{center}

At ambient temperature under vacuum of typically $\unit[1 \times 10^{-6}]{mBar}$, we measure a vibration level less than $\unit[4]{pm}$. At low temperatures, we see an increase to about $\unit[25]{pm}$. This level is consistently reached for several setup configurations, including different fiber and sample mounting and the integrated electronics for applying microwaves. Figure~\ref{fig:performance}~(a) shows the integrated noise spectrum of the measured length fluctuations and Fig.~\ref{fig:performance}~(b) an individual frequency spectrum. The length fluctuations show a small low-frequency contribution, which we attribute to the cold head cycle of the cryostat, running with a speed of about $\unit[1.4]{Hz}$ (settings: $\unit[25]{Hz}$ compressor and $\unit[70]{Hz}$ cold head). For all configurations, the spectrum shows a jump between $\unit[2]{kHz}$ and $\unit[4]{kHz}$, which fits the first resonance of the fiber positioning system along the cavity axis. Most configurations show an almost flat spectrum for frequencies higher than $\unit[5]{kHz}$.\\
Another measure of the vibration level is a resonant laser scan over the cavity transmission, shown in Fig.~\ref{fig:performance}~(c). The cavity is probed with a resonant $\unit[637]{nm}$ laser from the fiber side and the transmission signal is recorded with a fiber-coupled single photon detector (Laser~Components~COUNT-10C-FC). With the cavity length, we determine a dispersion of $\unit[20]{MHz/pm}$. In the previous measurement in Fig.~\ref{fig:operation}~(a) the linewidth is measured by scanning the cavity with a high scan speed to 'freeze' the vibrations, revealing the intrinsic cavity linewidth (shown in Fig.~\ref{fig:operation}~(a)). In contrast, the cavity is at a fixed length in Fig.~\ref{fig:performance}~(c), but the resonant laser frequency is swept to measure the linewidth. The data is fitted with a Voigt profile with a fixed Lorentzian part set to the intrinsic cavity linewidth. Vibrations lead to a Gaussian broadening, which can be transformed into a RMS vibration level with the dispersion relation. For the scans shown in Fig.~\ref{fig:performance}~(c), we measure a vibration level of $\unit[(25.1 \pm 1.1)]{pm}$, which is close to the $\unit[22]{pm}$ measured in Fig.~\ref{fig:performance}~(a).\\
We want to emphasize that the low vibration level is reproducibly reached for different configurations, in which the fiber or sample mounting is changed. Furthermore, it is also independent of the additional integration of microwave cables, a support PCB, and wire bonds to the sample mirror. The microcavity insert was reassembled in between different configurations, and thermally cycled for more than $\unit[50]{}$ times over two years without a degradation in performance. We do not see a vibration dependence on the exact wiring of cables used to connect the piezo elements or the microwave wires. Moreover, the vibration level did not change after exchanging the cold head.\\
In addition to vibrations $\unit[> 1]{Hz}$, the cavity can show drifts on longer timescales. When operated with higher incident laser power ($\unit[> 1]{\upmu W}$) or microwave power ($ > \unit[25]{dBm}$ measured before the input of the cryostat, see next section for the transmission losses), cavity length drifts of more than one linewidth can be observed. However, after some time (order of minutes) the drifts reduce to a new equilibrium.\\
The closed-cycle operation of the cryostat allows to operate the cavity at a constant temperature (cryostat temperature stability $\unit[< 25]{mK}$) and to maintain the exact cavity spot for a longer time. In previous work \cite{herrmann_coherent_2024}, measurements with the same cavity-coupled diamond Tin-Vacancy center were conducted for more than two months with preserved cavity parameters.\\

\onecolumngrid
\twocolumngrid

\begin{figure}[ht]
    \centering
    \includegraphics[width=\linewidth]{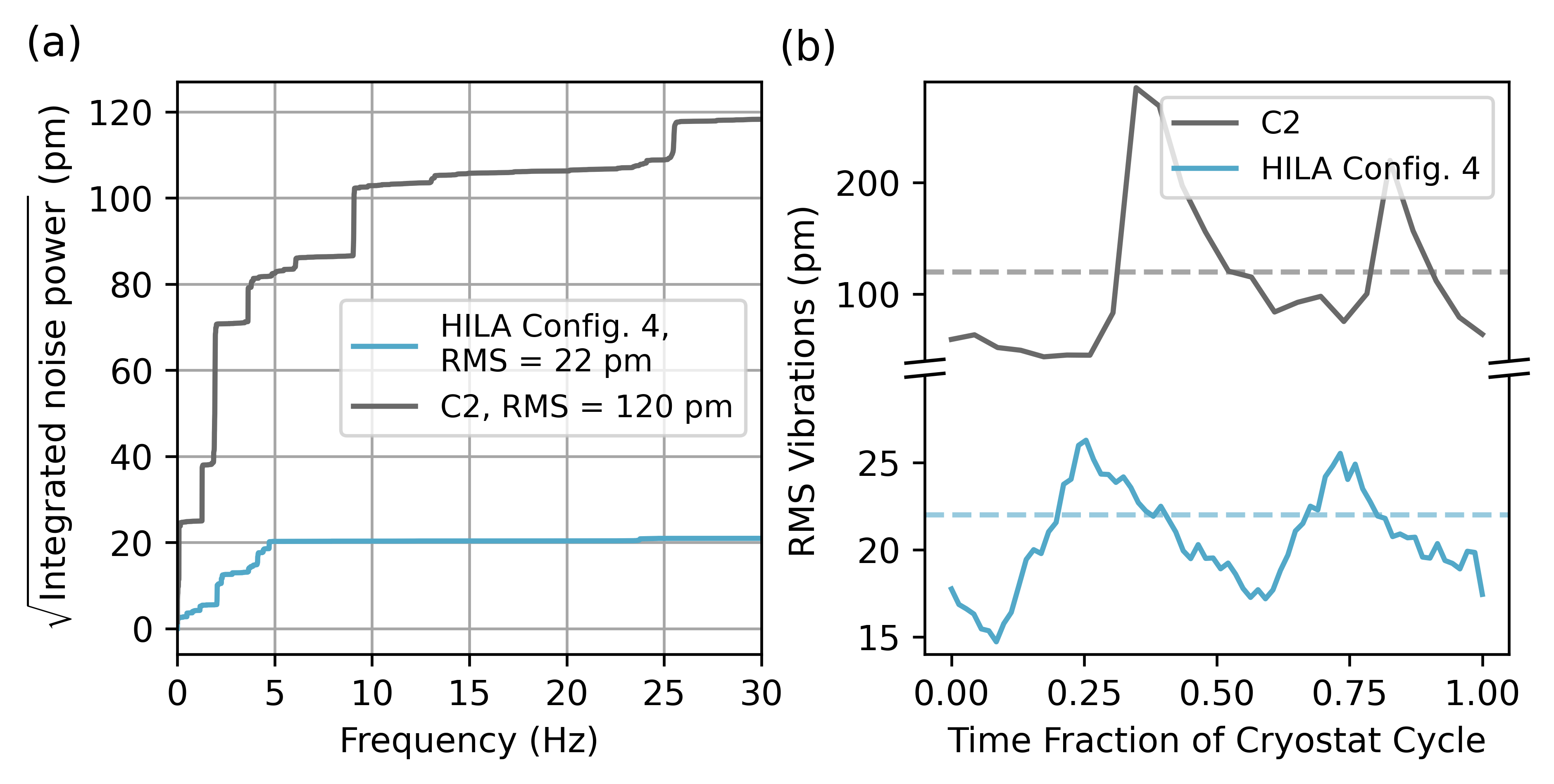}
    \caption{Comparison of vibration level achieved in configuration four with the HILA cryostat (light blue) versus the C2 optical cryostation (grey), used in an earlier experiment \cite{ruf_resonant_2021}. (a) Integrated noise spectra show that the mechanical resonance frequencies of the positioning system are excited in both cryostats, although at a much lower level in the HILA. Frequencies larger than $\unit[5]{kHz}$ are almost completely suppressed in the HILA. (b) RMS vibration level scaled to the time fraction of the cryostat cycle (about $\unit[0.7]{s}$ for the HILA and $\unit[1.2]{s}$ for the C2). The HILA shows a much smaller susceptibility to 'kicks' within the cold head cycle.}
    \label{fig:comparison}
\end{figure}

The same microcavity insert (JPE CVIP 1 with CPSHR1-a) was operated in earlier work \cite{ruf_resonant_2021} in a standard optical cryostation (Montana~Instruments~C2). Thanks to the similarity in setup design, it is instructive to compare both performances, see Fig.~\ref{fig:comparison}. The much better vibration isolation of the HILA leads to an almost six times improved RMS vibration level and a much smaller dependence on the cryostat cycle.\\
There are several further improvements of the vibration level possible: the damping with the PTFE~film as explained in section~\ref{sec:setup} can be optimized. Furthermore, the stability of the system would benefit from an increased mechanical resonance frequency of the positioning system, which can be achieved by further engineering. For example, a higher resonance frequency might be achievable by removing the fine piezo element and losing the ability for precise positioning and scanning of the fiber. Along the cavity axis, this could be compensated by using the shear piezo element with a scanning range of about $\unit[300]{nm}$ at low temperatures. Another possibility would be to increase the temperature of the positioning system. At room temperature, under vacuum, and with a running cold head, we observe a much better vibration level. An explanation could be that temperature-dependent material properties lead to an increased vibration level at low temperatures. Thus, putting the fiber positioning system at a higher temperature (for example on stage one of the cold head at $\unit[90]{K}$), might further improve the stability.\\
Furthermore, if it is compatible with the sample, it might be beneficial to operate the fiber in contact with the sample mirror. In a similar setup with a fiber-based microcavity in a closed-cycle cryostat \cite{pallmann_highly_2023}, this resulted in a vibration level improved by a factor of six.\\
So far, we have only discussed the passive stability of the system. Most mechanical resonance frequencies are present below $\unit[10]{kHz}$, which makes active length stabilization appealing. The use of a high bandwidth piezo element in a similar system enabled a locking bandwidth of $\unit[44]{kHz}$ \cite{janitz_high_2017}, which would be sufficient to compensate for the resonance frequencies present in our system. Active stabilization would require some experimental overhead like an additional laser and a broad (or second) cavity stopband, to find a cavity resonance that can be used to generate the error signal. This is especially challenging for a short cavity with a large free spectral range. Furthermore, the locking laser could lead to stimulated emission of the cavity-coupled emitters \cite{raman_nair_amplification_2020}. To avoid this, a second cavity formed by an additional fiber dimple can be utilized for active length stabilization \cite{ulanowski_spectral_2024}.

\section{Cavity-coupled Diamond Color Centers with Microwave Control}

\begin{center}
\begin{figure*}[ht]
    \centering
    \includegraphics[width=0.9\linewidth]{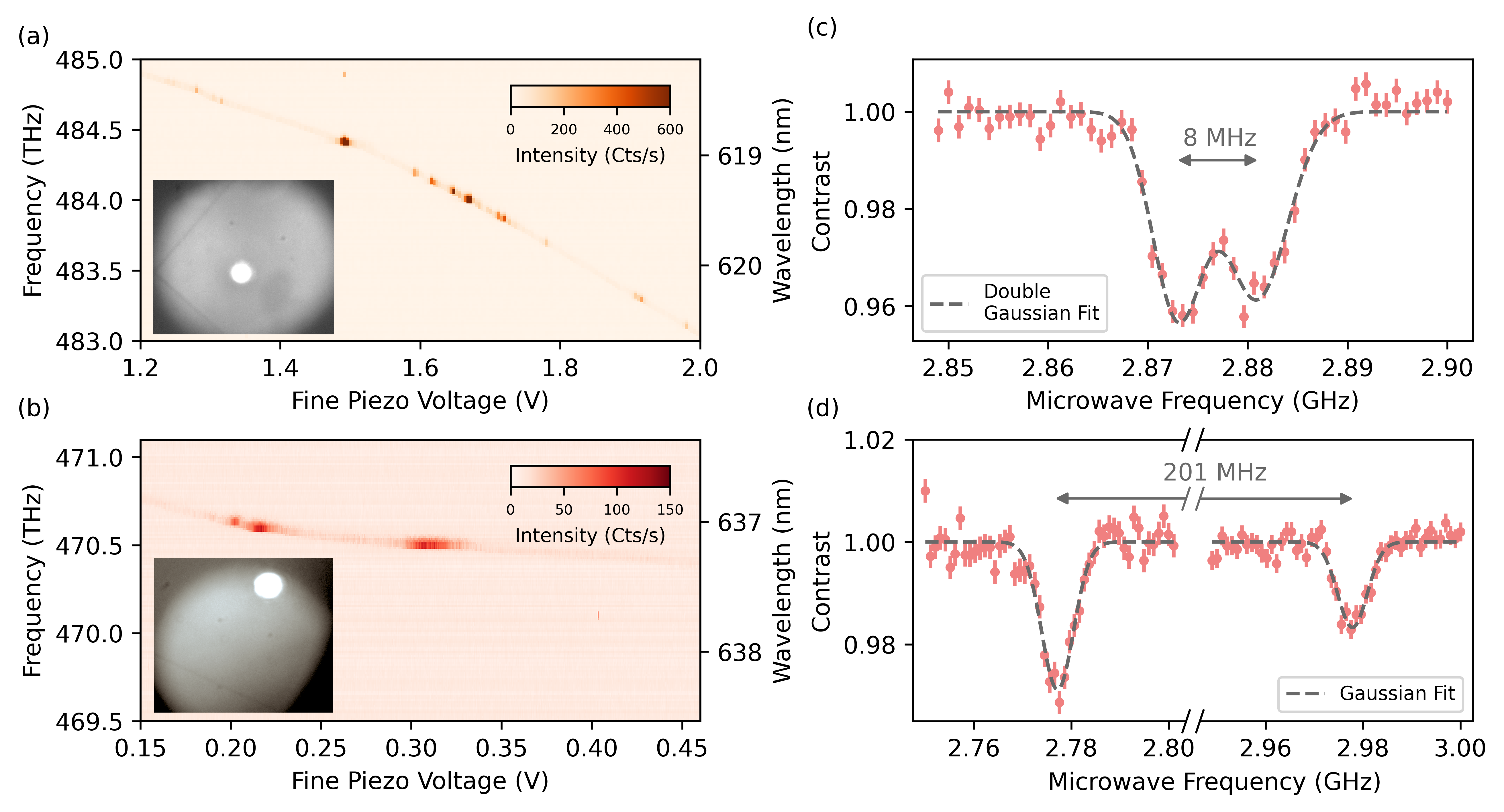}
    \caption{Platform versatility and microwave spin addressing. The insets of (a) and (b) show an about $\unit[50\:\text{x}\:50]{\upmu m^2}$ large section. (a) Cavity-coupled SnV centers under off-resonant excitation. The cavity mode (faint straight line) is scanned over the $\unit[619]{nm}$ and $\unit[620]{nm}$ zero-phonon line transitions of several SnV centers within the mode volume, which appear as bright spots. The inset shows the diamond device hosting the color centers bonded to the sample mirror. The position of the cavity on the sample is indicated by the bright laser spot, stemming from a green laser for which the cavity is mostly transparent. (b) Cavity-coupled NV centers. Under green excitation, the NV center is mostly initialized into the spin ground state. When the cavity becomes resonant, it enhances the $E_x$ or $E_y$ zero-phonon line transitions. The inset shows the thinned-down diamond membrane bonded to the sample mirror and the approximately $\unit[30]{\upmu m}$ distant gold stripline. The nonlinearity in the cavity resonance in (a) and (b) stems from a small cavity length drift during the data aquisition. (c) ODMR measurement of a cavity-coupled NV center without an applied magnetic field. The NV zero field splitting around $\unit[2.87]{GHz}$ can be observed as a resonance. The splitting of $\unit[8]{MHz}$ can originate from a residual magnetic field or a strongly coupled carbon nuclear spin in the vicinity of the NV center. (d) ODMR measurement of a cavity-coupled NV center with an external static magnetic field. In the presence of the magnetic field, the two resonances associated with the NV electron spin are separated by the Zeeman splitting $2\gamma_e B_z$, with the NV electron spin gyromagnetic ratio $\gamma_e$ and the magnetic field $B_z$. The splitting of $\unit[201]{MHz}$ corresponds to $B_z \approx \unit[36]{G}$. The lower contrast for the high-frequency transition is caused by increased microwave losses at this frequency. Note that the ODMR measurements were acquired at a cavity spot around $\unit[20]{\upmu m}$ distant to the stripline, which is a different spot than shown in (b). All measurements are performed at a sample temperature of about $\unit[8]{K}$.}
    \label{fig:coupling}
\end{figure*}
\end{center}

We showcase the versatility of our platform by coupling two types of diamond color centers to the cavity, namely Tin-Vacancy~(SnV) and Nitrogen-Vacancy~(NV) centers, shown in Fig.~\ref{fig:coupling}~(a) and (b), respectively. The SnV sample is a $\unit[70\:\text{x}\:70]{\upmu m^2}$ square diamond device with a thickness of about $\unit[2.2]{\upmu m}$ bonded to the sample mirror via van der Waals force \cite{riedel_deterministic_2017}. SnV centers are created by ion implantation in the range of the first intracavity electric field antinode. The NV sample is a $\unit[2\:\text{x}\:2]{mm^2}$ and about $\unit[50]{\upmu m}$ thick diamond membrane, with a central part thinned down to a thickness of about $\unit[5.8]{\upmu m}$, used in earlier work \cite{ruf_resonant_2021}. NV centers are created by electron irradiation and annealing \cite{ruf_optically_2019}. The cavity spot shown in the inset of Fig.~\ref{fig:coupling}~(b) is formed in the thinned-down region.\\
To demonstrate that diamond color centers are coupled to the cavity at low temperature, we use off-resonant $\unit[515]{nm}$ excitation light (H\"{u}bner~Photonics~Cobolt~MLD515) sent into the cavity via the fiber. The detection signal is collected from the sample mirror side by collimating the Gaussian beam leaving the cavity with the objective. The detection is filtered with a $\unit[600]{nm}$ longpass filter and measured with a spectrometer. The cavity resonance frequency is tuned around the emitter zero-phonon line by scanning the voltage of the three fine piezo elements. Once the cavity becomes resonant with an emitter, enhanced photoluminescence at that frequency is observed. An individual emitter can then be selectively coupled to the cavity by adjusting the fine piezo element voltage to the corresponding resonance frequency.\\
An essential capability for quantum information applications is qubit control, which can be efficiently achieved through driving the corresponding spin states with microwaves. To deliver microwaves close to the cavity spot, $\unit[65]{nm}$ thin gold striplines are partly embedded into the sample mirror with nano-fabrication methods \cite{bogdanovic_robust_2017}. The striplines are connected as described in section~\ref{sec:setup}. With a vector network analyzer, we measure around $\unit[6.5]{dB}$ transmission loss including the PCB at a frequency of $\unit[2.88]{GHz}$. The total cryostat transmission losses add up to about $\unit[25]{dB}$ due to several cable connections. To demonstrate the ability to address the spin states of the NV center, we measure an optically detected magnetic resonance~(ODMR) spectrum without and with a static magnetic field shown in Fig.~\ref{fig:coupling}~(c)~and~(d). The NV centers are excited with off-resonant green light sent in via the cavity fiber. Cavity-enhanced zero-phonon light is detected from the free space side with a single photon detector. This light is filtered with a $\unit[600]{nm}$~longpass filter, a $\unit[(640\pm5)]{nm}$~bandpass filter and an angle-tunable etalon filter (LightMachinery~custom~coating, full width at half maximum $\unit[\approx100]{GHz}$). The cavity is kept on resonance during the ODMR measurement by interleaved probing of the cavity transmission with a reference laser and a voltage feedback on the fine piezo element. An external magnetic field is provided by a neodymium disc magnet (Supermagnete~S-70-35-N), which is mounted on top of the cryostat vacuum chamber lid. Microwaves are delivered by a single-sideband modulated vector signal generator (Rohde~\&~Schwarz~SMBV100A) with an output power of $\unit[-5]{dBm}$ ($\unit[-3]{dBm}$ for the higher-frequency transition), which is amplified by $\unit[35]{dB}$~(MiniCiruits~ZVE-6W-83+). A reduction in NV photoluminescence is observed when the microwave driving is on resonance with one of the ground state spin transitions.

\onecolumngrid
\twocolumngrid

\section{Conclusion}

We present the design, assembly, and operation of a fiber-based microcavity platform with high passive stability and demonstrate the cavity coupling of diamond color centers together with the ability to address the spin of cavity-coupled NV centers with microwaves. The key elements for achieving high passive stability is the combination of a low-noise cryostat with incorporated vibration suppression and a positioning system with a high mechanical resonance frequency. The resulting separation of noise frequencies and positing system resonance frequencies leads to the high cavity length stability. Our system shows robustness and reproducibility over many thermal cycles with varied setup configurations. Moreover, the passive stability seems to be independent of experimental imperfections such as the exact wiring of cables or the tightness of screws. The vibrations in our system are mainly caused by the cold head, and we do not observe any influences on the vibration level of other external sources like acoustic noise.\\
With improvements such as demonstrated by a related system \cite{pallmann_highly_2023} or by active cavity length stabilization \cite{brachmann_photothermal_2016}, it might be possible to further reduce the vibration level. A stability improvement by another order of magnitude would render the effect of the vibrations on the maximum attainable Purcell factor almost negligible.\\
Notably, all parts for achieving the high passive vibration level, such as the floating baseplate cryostat and the fiber positioning system, are commercial.\\
For the estimated values for diamond color centers in Fig.~\ref{fig:schematic}~(a), the achieved passive stability would allow for an air-like Purcell factor up to about~25, with even higher values possible for a diamond-like cavity, in conjunction with good collection efficiency. With the integration of microwaves for full qubit control, this system provides a promising platform for testing and utilizing efficient qubit-photon interfaces.

\clearpage

\section*{Acknowledgement}
We thank Arian Stolk and Kai-Niklas Schymik for feedback on the manuscript, Matteo Pasini for helpful discussions, Siebe Visser for technical support, Jason Mensingh and Olaf Benningshof for cryogenic engineering support, Nico Alberts and Tim Hiep for machining the custom parts, Raymond Vermeulen and Raymond Schouten for development of custom electronics and Henri Ervasti for software development and support.\\
We thank Robbert Voorhoeve, Tom Duivenvoorde, Teun van den Dool and Gert Witvoet from TNO for support on the setup design. Furthermore, we thank Hans Spierdijk and Jan de Vreugd from TNO for the design of the objective stage.\\
We acknowledge support on device fabrication from Hans van den Berg, Jasper Flipse and Nick de Jong from TNO.\\
We thank Bart van Bree and Huub Janssen from JPE for helpful discussions and support on the cryo positioning stage.\\
We thank Matt Ballinger and Alex Crane from Montana Instruments and Robert Janz and David Appel from Quantum Design for helpful discussions and the installation and support of the HILA cryostat.\\
We acknowledge funding from the Dutch Research~Council~(NWO) through the Spinoza Prize 2019 (project number SPI~63-264). We further acknowledge financial support from the EU Flagship on Quantum Technologies through the project Quantum Internet Alliance (EU~Horizon~2020, grant agreement no.~820445). This research is supported by the Early Research Programme of the Netherlands Organisation for Applied Scientific Research (TNO). Additional support from the Top Sector High Tech Systems and Materials is highly appreciated.

\section*{AUTHOR DECLARATIONS}

\subsection*{Conflict of Interest}
The authors have no conflicts to disclose.

\subsection*{Author Contributions}
Y. H. and J. F. contributed equally to this work. Y.H. and J.F. conducted the experiments and analyzed the data. S.S. characterized the cavity fibers and measured the data for one of the setup configurations. C.W. developed the ODMR measurements. M.R. developed parts of the device fabrication process and designed together with M.J.W., Y.H. and J.F. the setup. Y.H., J.F., L.J.F. and M.J.W. built the setup. J.M.B., C.S. and Y.H. fabricated the Tin-Vacancy diamond device. L.G.C.W. characterized the Tin-Vacancy diamond device and cavity fibers in a room-temperature cavity. M.R. fabricated the Nitrogen-Vacancy diamond device. M.E. fabricated the cavity fibers. Y.H., J.F. and R.H. wrote the manuscript with input from all authors. R.H. supervised the experiments.\\

\subsection*{Data Availability}
The data that support the findings of this study are openly available on 4TU.ResearchData: 'Data underlying the publication "A Low-Temperature Tunable Microcavity featuring', at https://www.doi.org/10.4121/451152e2-a4d4-4e42-96e0-4147afb1e45c, reference number \cite{herrmann_data_2024}.

\appendix

\section{Derivation of the Maximum Attainable Purcell Factor}
\label{app:derivation}

The Purcell factor is defined as\cite{fox_quantum_2006} 
\begin{equation}
    F_{P} = \frac{3}{4\pi^2} \left(\frac{c}{n\nu}\right)^3 \frac{Q}{V},
\label{equ:Purcell}
\end{equation}
with the cavity resonance frequency $\nu$, refractive index $n$, cavity quality factor $Q$, cavity mode volume $V$, refractive index of diamond $n$ and the speed of light $c$.

The spectral overlap of an emitter with a linewidth much smaller than the cavity linewidth is given by \cite{van_dam_optimal_2018}
\begin{equation}
    \xi_s(\Delta\nu) = \frac{1}{1 + \frac{4Q^2}{\nu^2} \Delta\nu^2},
\label{equ:spectral_overlap}
\end{equation}
with the cavity resonance frequency $\nu$ and the emitter-cavity detuning $\Delta\nu$.

The cavity frequency fluctuations can be modeled by a Gaussian distribution, where the probability density function reads \cite{ruf_resonant_2021}
\begin{equation}
    f(\Delta\nu) = \frac{1}{\sqrt{2\pi\sigma_{\nu}^2}} e^{-\Delta\nu^2/2\sigma_{\nu}^2},
\label{equ:prob_density_function}
\end{equation}
with the RMS value of the cavity frequency fluctuations $\sigma_{\nu} = s\sigma$, where $s$ is the cavity mode dispersion slope and $\sigma$ the RMS cavity length fluctuations.

For an emitter on cavity resonance, the effective vibration-averaged Purcell factor can be calculated with
\begin{equation}
    F_{P,vib} = F_{P} \int_{-\infty}^{\infty} d(\Delta\nu) \xi_s(\Delta\nu) f(\Delta\nu).
\label{equ:Purcell_vib}
\end{equation}

For a given cavity geometry, which determines the cavity mode volume $V$ and the cavity mode dispersion slope $s$, the maximum attainable Purcell factor follows as
\begin{equation}
    F_{P,max} = \max_{Q>0}{F_{P,vib}} = \frac{3}{4\pi^2} \left(\frac{c}{n\nu}\right)^3 \frac{1}{V} \sqrt{\frac{\pi}{2(s\sigma)^2}} \frac{\nu}{2}.
\label{equ:max_attain_Purcell}
\end{equation}

To obtain equation (\ref{equ:max_attain_Purcell}) we make use of
\begin{equation}
    \int_{-\infty}^{\infty} dx \frac{e^{-x^2/B}}{1 + A x^2} = \frac{\pi e^{1/AB} \text{erfc}{\left(1/\sqrt{AB}\right)}}{\sqrt{A}},
\label{equ:integral}
\end{equation}
with $A > 0$, $B > 0$ and the complementary error function $\text{erfc}(x) = 1 - \text{erf}(x)$ as well as
\begin{equation}
    \max_{x>0}{\left(e^{1/x^2} \text{erfc}{(1/x)}\right)} = 1.
\label{equ:maximum}
\end{equation}

\section{Cavity Fiber Positioning System}

\begin{figure}[ht]
    \centering
    \includegraphics[width=0.9\linewidth]{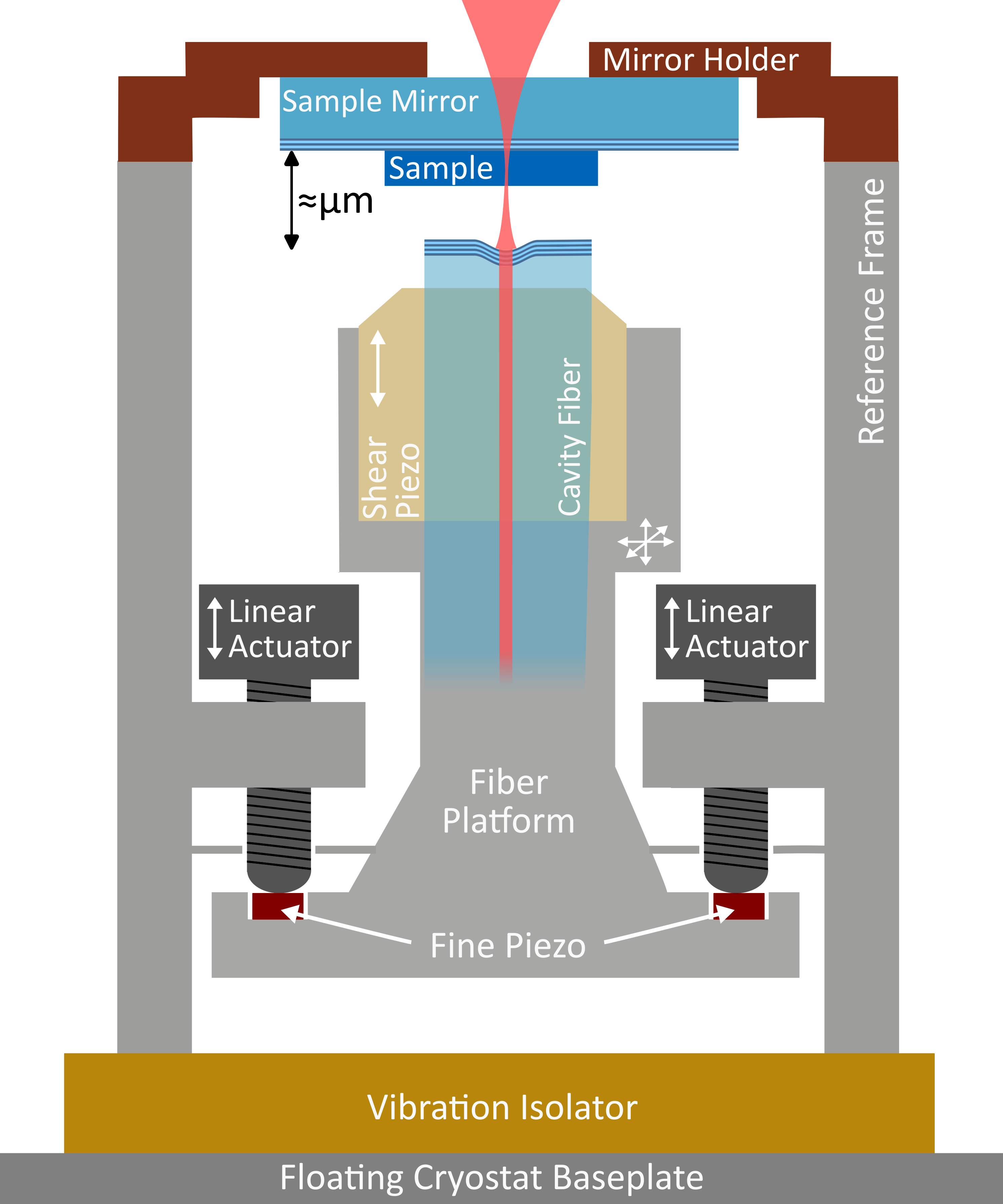}
    \caption{Sketch of the microcavity insert. Three linear actuators (JPE~CLA2201) position the cavity fiber in front of the sample mirror: moving all three actuators simultaneousely changes the cavity length in axial direction, while a single actuator movement changes the lateral position (together with a small change in the fiber angle). Each axis can additionally be controlled with a fine scanning piezo, which are clamped between the fiber platform and the linear actuators. Optionally, the fiber can be glued on a shear piezo for axial cavity length control. A passive vibration isolator (JPE~CVIP1) is placed underneath the positioning system (JPE~CPSHR1-a).}
    \label{fig:positioning}
\end{figure}

Fig.~\ref{fig:positioning} shows a sketch of the full microcavity insert, which is presented in the main text as a rendered CAD drawing in Fig.~\ref{fig:drawing}~(b).

\section{Photographs of the Setup}

\begin{figure*}[ht]
    \centering
    \includegraphics[width=\linewidth]{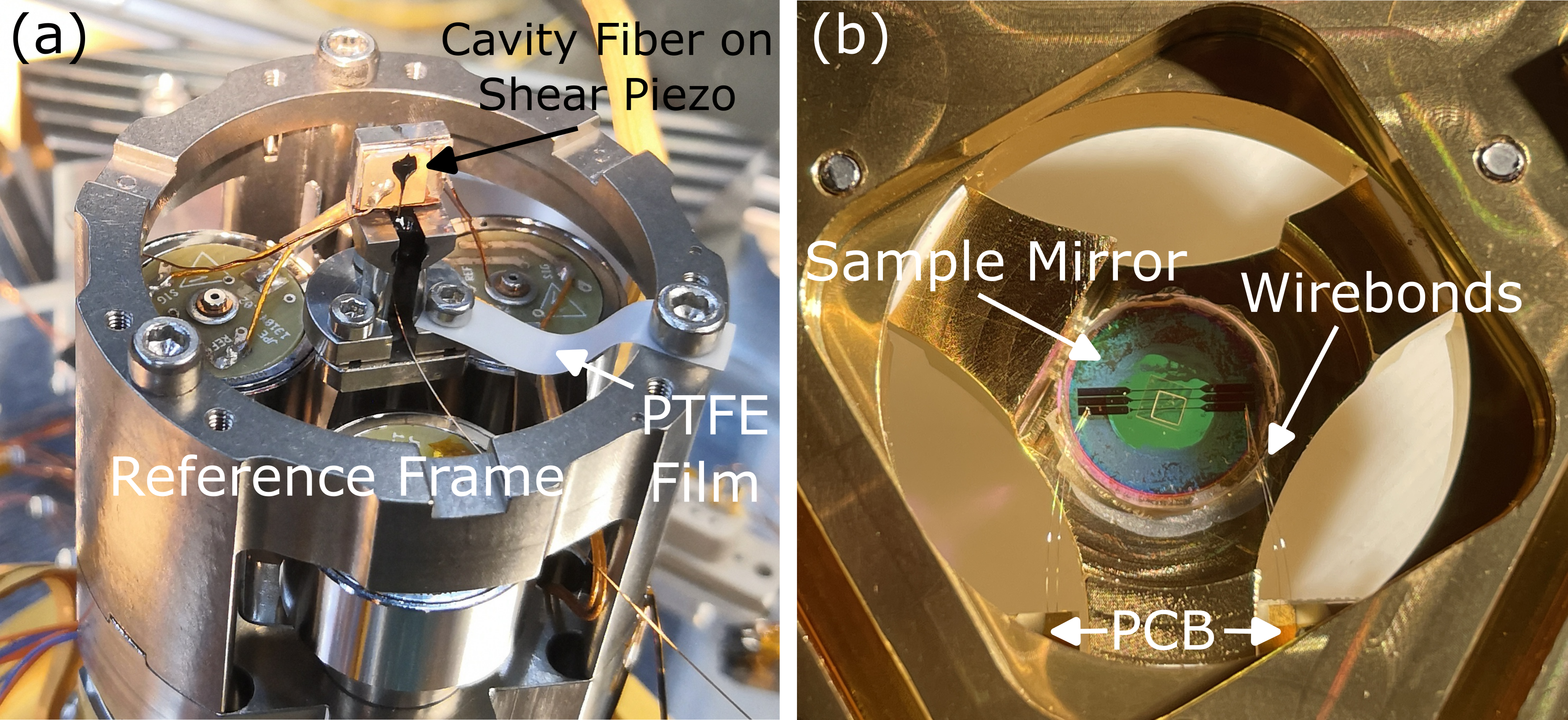}
    \caption{Photograph of the fiber and sample mounting. (a) The lower half of the cavity: cavity fiber glued to the shear piezo element for fast cavity length control. A $\unit[0.25]{mm}$ thin PTFE film is used to introduce additional damping, which can improve the passive stability. The platform with the fiber mount is moved in situ by the positioning system. (b) The upper half of the cavity: the sample mirror with embedded microwave stripline and a partially thinned-down diamond membrane. A support PCB is glued to the mirror holder adapter to connect the microwave lines via wire bonds.}
    \label{fig:fiber_sample}
\end{figure*}

\begin{figure*}[ht]
    \centering
    \includegraphics[width=\linewidth]{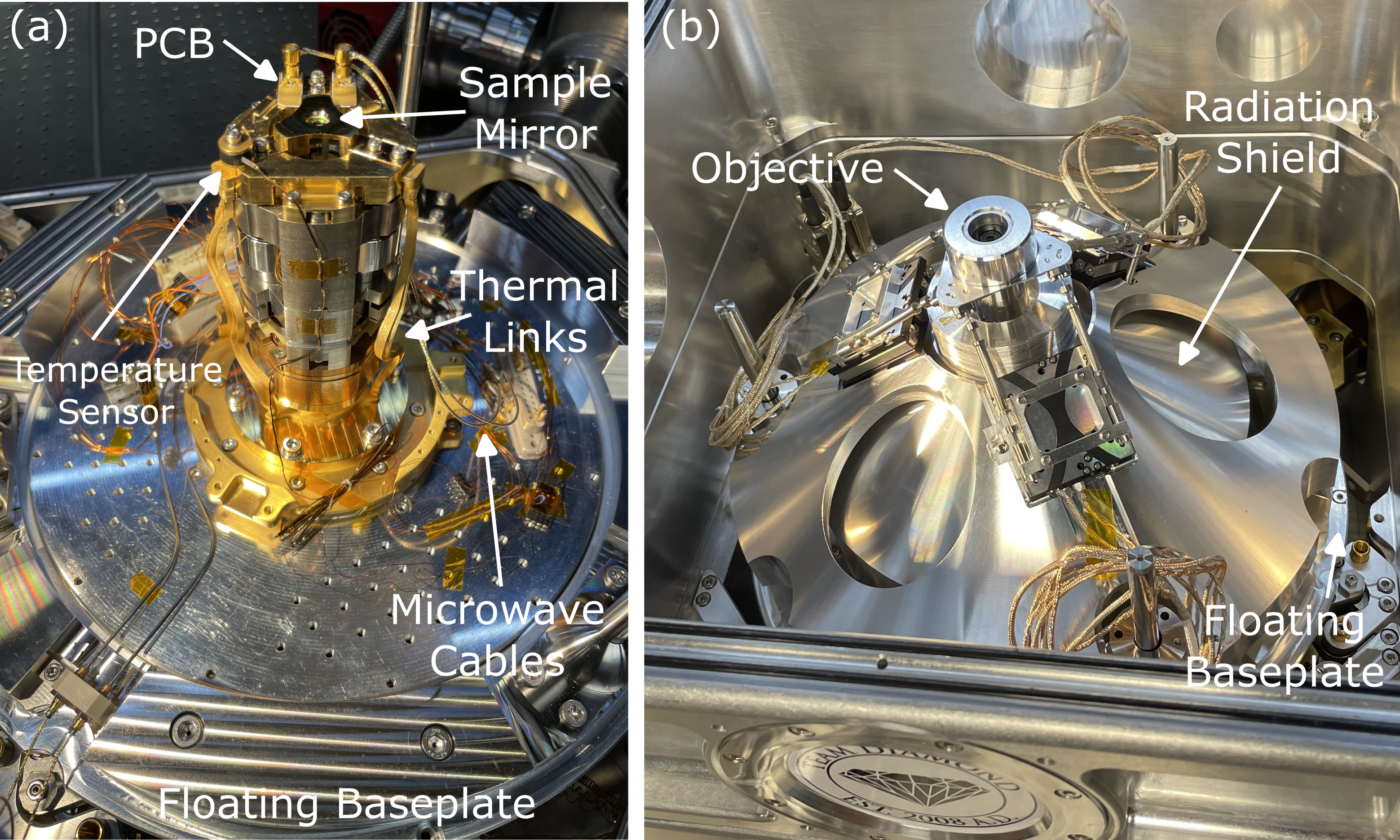}
    \caption{Photograph of the microcavity insert and the enclosing objective cone. (a) The insert is placed on the $\unit[5.6]{K}$ platform of the cryostat baseplate. Three thermal links ensure a good thermalization of the mirror holder to the baseplate. The temperature reached on top of the stack is monitored with a temperature sensor. The microwave lines of the PCB are connected with flexible coax cables to the baseplate to minimize the transfer of vibrations. On the baseplate, microwaves are guided by rigid cables to the vacuum feedthrough. (b) The insert is enclosed by a radiation shield and a stainless steel cone mounting the objective with linear actuators in a tripod setting. The radiation shield is made of a lower cone and an upper cylinder with an opening fitting the objective. The upper part is screwed into the lower part. The cone and objective are at room temperature inside the cryostat vacuum chamber.}
    \label{fig:insert_chamber}
\end{figure*}

\begin{figure*}[ht]
    \centering
    \includegraphics[width=\linewidth]{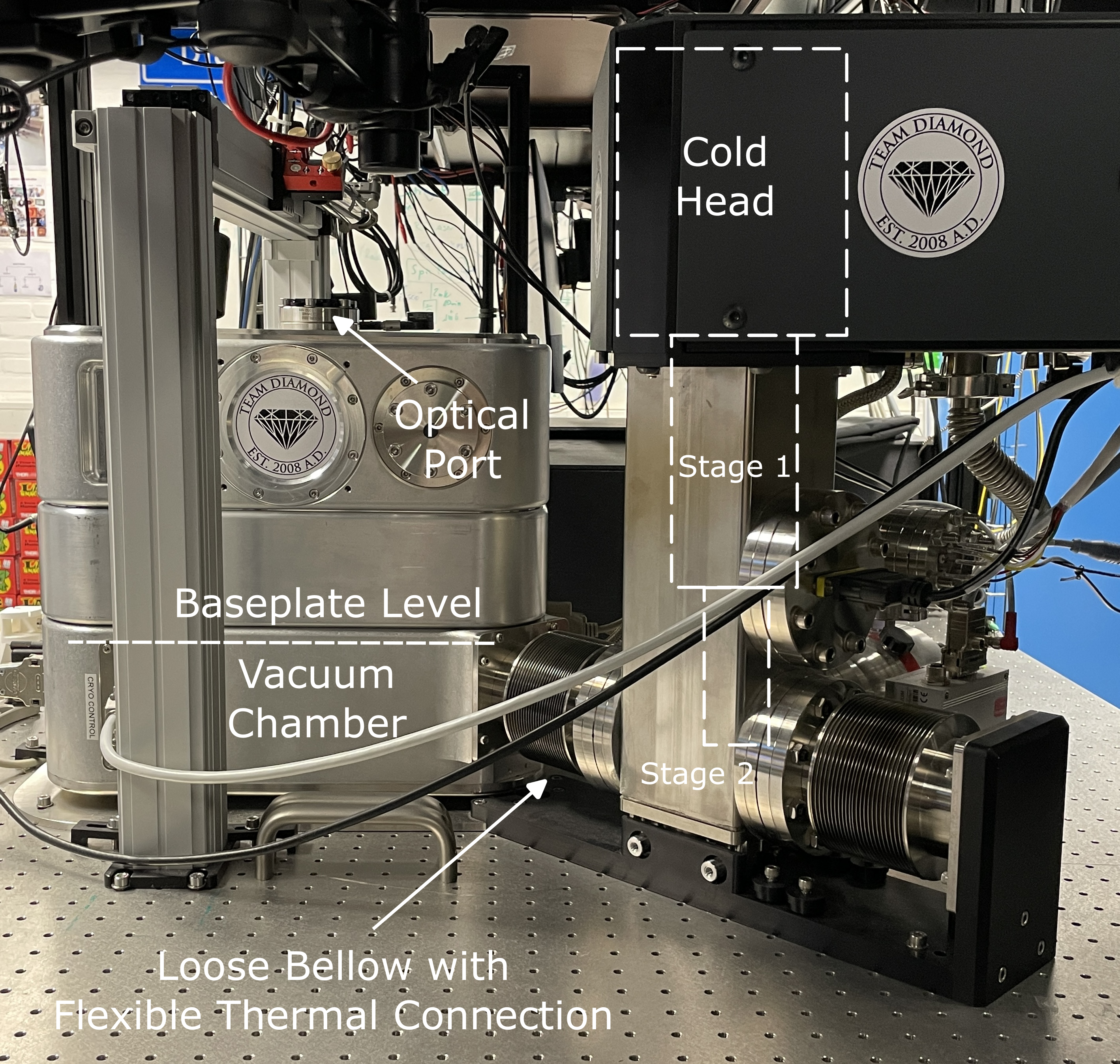}
    \caption{Photograph of the HILA cryostat vacuum chamber with the off-table cold head inside the tower assembly, mounted by a rack next to the optical table. Both stages of the cold head are connected over flexible thermal links with the floating baseplate inside the vacuum chamber.}
    \label{fig:chamber}
\end{figure*}

Photographs of individual parts of the setup and the cryostat can be found in Fig.~\ref{fig:fiber_sample}~to~Fig.~\ref{fig:chamber}.

\clearpage
\bibliography{bib}

\end{document}